\documentclass[aps]{revtex4}
\usepackage{amssymb}

\input{tcilatex}

\begin{document}

\title{Critical gaps of first-order phase transition in infinitely long
Ising cylinders with antiperiodically joined circumference }
\author{$^{1}$Tsong-Ming Liaw, $^{2}$Ming-Chang Huang, $^{1,3}$Simon C. Lin,
and $^{2}$Yu-Pin Luo}

\begin{abstract}
Based on the analytic expression of free energy for infinitely long Ising
strip with finite width joined antiperiodically on a variety of planar
lattices, we show the existence of first-order phase transition at the
critical point of Ising transition. The critical gaps of the transition are
also calculated analytically by measuring the discontinuities in the
internal energy and the specific heat.
\end{abstract}

\affiliation{${\ }^{1}$Computing Centre, Academia Sinica, 11529 Taipei,Taiwan\\
${\ }^{2}$Department of Physics, Chung-Yuan Christian University, Chungli
320,Taiwan\\
$^{3}$Institute of Physics, Academia Sinica, 11529 Taipei,Taiwan }
\maketitle


\section{Introduction}

In the study of critical phenomena, the investigations about finite-size
effect have been very active\cite{privman1}. These introduce more stringent
test on the universality by including universal critical amplitudes and
amplitude relations\cite{bb}. The finite-size effect is also important in
the comparison between theory and experiment. For example, it relies on the
knowledge of finite-size effect to extract the bulk properties from the
experimental results. In the study of finite-size effect, a semi-infinite
system for which the effect can be singled out is always of fundamental
importance. Thus, the system of an infinitely long Ising cylinder with
finite circumference has been studied for some times\cite%
{os,fs1,ferdi,lhlw,wuhua}. To be more specific, we define the system as
infinite extension along the $x$-axis and finite extension $L$ along the $y$%
-axis. To form an infinitely long Ising cylinder, we can join the finite
width periodically or antiperiodically. For periodically joined
circumference (PJC, hereafter), the condition imposed on the coupling of the
boundary spin variable $\sigma _{m,L}$ is $\sigma _{m,L+1}=\sigma _{m,1}$
with $-\infty <m<\infty $. For antiperiodic joined circumference (AJC,
hereafter), the condition becomes $\sigma _{m,L+1}=-\sigma _{m,1}$. For the
system with PJC, the finite-size scaling behaviors of the shift of critical
temperature $\delta \left( L\right) $, defined as $\delta \left( L\right)
=\left\vert T_{\max }\left( L\right) -T_{c}\right\vert /T_{c}$ with the
specific-heat peak $T_{\max }\left( L\right) $ and the bulk critical
temperature $T_{c}$, was known. The exact scaling form was shown to be $%
\delta \left( L\right) =b\ln L/\left( L\right) ^{2}$, where $b$ is a
function of coupling ratios and independent of the size\cite{ferdi}.
Recently, we studied the finite-size scaling of the distribution of
partition function zeros of such a system\cite{huanga}. The leading
finite-size scaling behavior of the imaginary part of a zero labelled by $j$
is given as $\func{Im}z_{j}\left( L\right) \thicksim L^{-1/\nu }$ with the
correlation length exponent $\nu $\cite{itzykson}; and the leading scaling
behavior of the real part of the lowest zero $\left( j=1\right) $ can be
written as $\left\vert \func{Re}z_{1}\left( L\right) -z_{c}\right\vert
\thicksim L^{-\lambda _{zero}}$. Here $\lambda _{zero}$ is another critical
exponent which is closely related to the shift exponent $\lambda $
characterizing the scaling behavior of $\delta \left( L\right) $\cite%
{yang,lee,fs2,janke}. Our results showed that the system with PJC has $\nu
=1 $ and $\lambda _{zero}=2$. \ \ 

For the system with AJC, the influences of the finite-size effect on
thermodynamic quantities are quite different from those for the system with
PJC. It was shown that the critical point of two-dimensional Ising phase
transition is one of the partition function zeros for the system with AJC,
and the singularity associated with the zero at the critical point is a
first-order phase transition for finite circumference\cite{huanga}. As the
length of circumference goes infinite, the effect of antiperiodic boundary
condition becomes negligible, and the singularity changes to be a
second-order phase transition. This is a very intriguing feature. It is
known that antiperiodic boundary condition may enhance the appearence of the
Block walls in spin configurations. In general, the existence of Block walls
will increase the free energy but suppress the energy fluctuations of the
system. However, for an infinitely long Ising cylinder, based on the study
of the cumulative distribution of partition function zeros we found that the
existence of the zero mode of spin wave is solely responsible for the
occurence of the non-analyticity at the critical point. We notice that the
appearence of first-order phase transition does not contradict with the
Mermin-Wigner theorem\cite{mermin,anderson}. Because the zero mode of spin
wave always appear in the spin configurations, the system can not be viewed
as an effective one-dimensional system with short-range Hamiltonian.
First-order transitions are characterized by the first derivative of the
free energy at the transition point. In this paper, we further analyze the
nature of the first-order transition quantitatively by directly measuring
the discontinuities in the internal energy and the specific heat for the
systems with AJC on planar square, triangular, and hexagonal lattices with
anisotropic couplings.

This paper is organized as follows. In Sec. II, the analytic expressions of
free energy density for an infinitely long Ising cylinder with PJC and AJC
on a variety of planar lattices are given and discussed. In Sec. III, we
perform explicit calculations for the internal energy and the specific heat.
Our results show that there exists discontinuities locating at the critical
point of two-dimensional Ising transition for AJC. Then, in Sec. IV we
measure the discontinuities appearing in the internal energy and the
specific heat explicitly. Finally, we summarize the results in Sec. V.

\section{Free energy}

Consider an infinitely long strip with finite width $L$ on plane triangular
or hexagonal lattices. The two infinitely extended sides may be joined
periodically or antiperiodically. Then, Ising model with ferromagnetic
couplings between nearest neighbors is defined on such infinitely long
cylinder. The corresponding form for the free energies per site per $k_{B}T$
was shown to be 
\begin{eqnarray}
f\ &=&\ -\ln R-\frac{1}{2sL}\sum\limits_{\overline{p}=0}^{L-1}\int_{0}^{2\pi
}\frac{d\phi }{2\pi }\ln \left\{ A_{0}-A_{1}\cos \left[ \frac{2\pi (%
\overline{p}+\Delta )}{L}\right] \right.  \nonumber \\
&&\left. -A_{2}\cos \phi -A_{3}\cos \left[ \frac{2\pi (\overline{p}+\Delta )%
}{L}-\phi \right] \right\} ,
\end{eqnarray}%
with $s=1$ for triangular lattice and $s=2$ for hexagonal lattice. This
result can be obtained either by direct derivation\cite{lhlw} or by taking
one side to infinite in the results of finite lattices with appropriate
boundary conditions\cite{wuhua,mcwu}.

In the above expression, $R$ is some lattice-dependent function of
temperature and irrelevant to the physical singularity of the theory, and $%
A_{\mu }$'s ($\mu =0,$ $1,$ $2,$ and $3$) can be expressed in terms of
particular functionals as 
\begin{eqnarray}
A_{0} &=&\alpha _{0}^{2}+\alpha _{1}^{2}+\alpha _{2}^{2}+\alpha _{3}^{2}, \\
A_{1} &=&2\,(\alpha _{0}\alpha _{1}-\alpha _{2}\alpha _{3}), \\
A_{2} &=&2\,(\alpha _{0}\alpha _{2}-\alpha _{3}\alpha _{1}), \\
A_{3} &=&2\,(\alpha _{0}\alpha _{3}-\alpha _{1}\alpha _{2}),
\end{eqnarray}
with the functionals $\alpha _{\mu }$'s given as 
\begin{eqnarray}
{\alpha _{0}} &=&1\ +\ t_{1}t_{2}t_{3}, \\
{\alpha _{1}} &=&t_{1}\ +\ t_{2}t_{3}, \\
{\alpha _{2}} &=&t_{2}\ +\ t_{3}t_{1}, \\
{\alpha _{3}} &=&t_{3}\ +\ t_{1}t_{2},
\end{eqnarray}
for the triangular lattice, and 
\begin{eqnarray}
{\alpha _{0}} &=&1 \\
{\alpha _{1}} &=&t_{1}t_{2}, \\
{\alpha _{2}} &=&t_{2}t_{3}, \\
{\alpha _{3}} &=&t_{3}t_{1},
\end{eqnarray}
for the hexagonal lattice. Here the $t_{i}$'s are given in terms of the
hyperbolic tangential functions as 
\begin{equation}
t_{i}\ =\ \tanh \eta _{i}\ 
\end{equation}
with $\eta _{i}=J_{i}/k_{B}T$,$\ $and $J_{i}$ is the coupling constant given
in the $i$'th direction as specified in Fig. 1. Note that the expression of
triangular lattice with $t_{3}=0$ gives the result of rectangular lattice.

The $\Delta $ value of Eq. (1) is one-half and zero for PJC and AJC
respectively. Thus, the exact forms of the free energies subject to the two
types of boundary conditions differ only in the possible $\left( \overline{p}%
+\Delta \right) $-values. To simplify the notation, we define $p=\overline{p}%
+\Delta $ and change the sum of Eq. (1) to be over the $p$-values. Then, $p$
is integer ranging from $0$ to $L-1$ for AJC, and it is half integer from $%
1/2$ to $L-1/2$ for PJC. When the width of Ising strip $L$ is extended to
infinite, the sum of Eq. (1) becomes a simple integration and the two
expressions of different boundary conditions reduce to the same result. In
this limit, the critical temperature of the Ising transition is determined
by the condition\cite{plech1}, 
\begin{equation}
A_{0}-A_{1}-A_{2}-A_{3}=0.
\end{equation}

The form of the free energy density of Eq. (1) can be rewritten as 
\begin{equation}
f\ =\ -\ln R\ -\ \frac{1}{2sL}\sum\limits_{p}\left\{ \ln \left[
A_{0}-A_{1}\cos \left( \frac{2\pi p}{L}\right) \right] +I(p,\eta )\right\} ,
\end{equation}%
where 
\begin{eqnarray}
I(p,\eta ) &=&\int_{0}^{2\pi }\frac{d\phi }{2\pi }\ln \left[ 1+G(p,\eta
)\cos (\phi -\psi (p,\eta ))\right] ,  \label{cefe0} \\
G(p,\eta ) &=&\frac{\left[ A_{2}^{2}+\ A_{3}^{2}+\ 2\,A_{2}A_{3}\cos (2\pi
p/L)\right] ^{1/2}}{A_{0}-A_{1}\cos (2\pi p/L)},
\end{eqnarray}%
and the angle $\psi (p,\eta )$ is defined by 
\begin{equation}
\tan \psi (p,\eta )=\frac{A_{3}\sin (2\pi p/L)}{A_{2}+A_{3}\cos (2\pi p/L)}.
\end{equation}

We notice that the integration of Eq. (17) may be completed to simplify the
expression. However, to calculate the derivatives of free energy we have to
be cautious of the continuity of the integrand to avoid ill-defined results.
Particularly, if any sigularity appears in the integrand, the integration
becomes improper. Therefore, for concreteness, in calculating the
derivatives of free energy we first proceed with the derivatives and then
perform the integration in Eq. (17). \ \ \ 

\section{Derivatives of free energy}

The dimensionless internal energy density $\epsilon $ and specific heat $c$
are defined as follows, 
\begin{eqnarray}
\epsilon _{i} &=&\frac{\partial f}{\partial \eta _{i}}, \\
c &=&-\eta ^{2}\frac{\partial \epsilon }{\partial \eta }\ =\ -\eta ^{2}\frac{%
\partial ^{2}f}{\partial \eta ^{2}}.  \label{dr2}
\end{eqnarray}
Note that the measurement of Eq. (20) depends on the reference scale which
is one of the coupling constants indexed by $i$. However, the expression of
the specific heat is invariant under the change of the reference scale since 
$\eta _{i}^{2}(\partial /\eta _{i})^{2}=\beta ^{2}(\partial /\beta )^{2}$
for any $i$.

The calculation of Eq. (20) contains the derivative of Eq. (17) which gives 
\begin{equation}
\frac{\partial I}{\partial \eta }=\ \int_{-\psi (p,\eta )}^{2\pi -\psi
(p,\eta )}\frac{d\Phi }{2\pi }\left[ \frac{\partial G(p,\eta )}{\partial
\eta }\frac{\cos \Phi }{1+G\left( p,\eta \right) \cos \Phi }\right] ,
\label{dIf}
\end{equation}%
where we neglect the label of reference scale, and the validity for
interchanging the derivative with the integration is, however, only ensured
at the points where the integrand is continuous for any $\Phi $. Using the
technique of contour integration, we obtain the result of Eq. (22) as 
\begin{equation}
\frac{\partial I}{\partial \eta }=(1-\frac{1}{\sqrt{1-G^{2}(p,\eta )}})\frac{%
\partial }{\partial \eta }\ln G\left( p,\eta \right) \quad \text{for}\
1-G^{2}(p,\eta )>0.
\end{equation}

As indicated, the analytic form of $\partial I/\partial \eta $ given by Eq.
(23) is valid only for positive function $1-G^{2}(p,\eta )$. To inspect the
domain of the functional values, we plot $1-G^{2}(p,\eta )$ versus $\eta $
for some $p$-modes in Fig. 2 which indicate that $1-G^{2}(p,\eta )$ is a
positive and concave function of $\eta $ for a given $p$-mode and the
absolute minimum locates at the point $p=0$ and $\eta =\eta _{c}$ with $%
1-G^{2}(0,\eta _{c})=0$. Here $\eta _{c}$ denotes the $\eta $ value at the
critical point of ising transition. These features are summarized in the
following lemma and corollary: \ 

\begin{lemma}
For the ferromagnetic couplings, $1-G^{2}(p,\eta )$ as a function of $\eta $
is bounded from below for a given $p$-mode.
\end{lemma}

\begin{proof}
The function of interest can be rewritten as 
\begin{equation}
1-G^{2}(p,\eta )=\frac{N\left( p,\eta \right) }{\left[ A_{0}-A_{1}\cos (2\pi
p/L)\right] ^{2}},
\end{equation}%
where the numerator $N\left( p,\eta \right) $ is given as 
\begin{equation}
N\left( p,\eta \right) =A_{1}^{2}\left[ \cos (2\pi p/L)-B\left( \eta \right) %
\right] ^{2}+C\left( \eta \right) ,
\end{equation}%
with 
\begin{eqnarray}
B\left( \eta \right)  &=&\frac{(A_{2}A_{3}+A_{0}A_{1})}{A_{1}^{2}}, \\
C\left( \eta \right)  &=&(A_{0}^{2}-A_{2}^{2}-A_{3}^{2})-\frac{%
(A_{2}A_{3}+A_{0}A_{1})^{2}}{A_{1}^{2}}.
\end{eqnarray}%
Since the denominator of Eq. (24) is strictly positive, the sign of the
function is completely determined by the numerator $N\left( p,\eta \right) $%
. For the ferromagnetic couplings, we have $B\left( \eta \right) \geq 1$,
which implies $N\left( p,\eta \right) \geq N\left( 0,\eta \right) $ for any
given temperature $\eta $. Furthermore, according to the definitions of the $%
A_{\mu }$'s given by Eqs. (2)-(5) we have 
\begin{equation}
N\left( 0,\eta \right) =\left[ \left( {\alpha _{0}-\alpha _{1}-\alpha
_{2}-\alpha _{3}}\right) \left( {\alpha _{0}-\alpha _{1}+\alpha _{2}+\alpha
_{3}}\right) \right] ^{2}\geq 0.
\end{equation}%
Therefore, $N\left( p,\eta \right) $ is bound from below so is the value of $%
1-G^{2}(p,\eta )$. \ \ \ \ \ 
\end{proof}

\begin{corollary}
For $p=0$ the minimum of $1-G^{2}(p,\eta )$ as a function of $\eta $ locates
at the bulk critical point $\eta =\eta _{c}$ with zero value, and for $p\neq
0$ the minimum is larger than zero.
\end{corollary}

\begin{proof}
The value of $N\left( 0,\eta \right) $ given by Eq. (28) vanishes only at
the critical point. This is due to the facts as follows: The critical
condition of Eq. (15) is equivalent to 
\begin{equation}
{\alpha _{0}-\alpha _{1}-\alpha _{2}-\alpha _{3}=0.}
\end{equation}%
Moreover, for the ferromagnetic couplings we have $\alpha _{0}>\alpha
_{k}\geq 0$ for $k=1,$ $2,$ and $3$. This leads the expression, ${\alpha
_{0}-\alpha _{1}+\alpha _{2}+\alpha _{3}}$, in Eq. (28) to be strictly
positive. Thence, among the minima of $N\left( p,\eta \right) $ for various $%
p$-modes there exist an absolute minimum $N\left( 0,\eta _{c}\right) $ whose
value is zero. In other words for $p\neq 0$ we have 
\begin{equation}
N\left( p,\eta \right) >N\left( 0,\eta \right) \geq N\left( 0,\eta
_{c}\right) =0.
\end{equation}%
Hence, the domain of the functional value of $1-G^{2}(p,\eta )$ is specified
by the conditions as 
\begin{equation}
1-G^{2}(p,\eta )>0\text{ \ \ \ for }p\neq 0\text{ }
\end{equation}%
and 
\begin{equation}
1-G^{2}(0,\eta )\geq 1-G^{2}(0,\eta _{c})=0,
\end{equation}%
and the proof is completed. 
\end{proof}

According to the corollary, the case of $1-G^{2}(p,\eta )<0$ never happens
for the case of ferromagnetic couplings. However, the situation, $%
1-G^{2}(p,\eta )=0$, does occur at one point for which the expression of Eq.
(23) is invalid. Thus, the result of Eq. (23) fails only at the critical
point $\eta =\eta _{c}$ for AJC, and it is valid at any temperature for the
system with PJC. With the exclusion of the exceptional point from the result
of Eq. (23), we can obtain the internal energy $\epsilon $ of Eq. (20) as \
\ 
\begin{equation}
\epsilon \ =\epsilon _{s}+\ \frac{1}{2sL}\sum\limits_{\overline{p}%
}D_{\epsilon }(p,\eta ),  \label{de}
\end{equation}%
with 
\begin{equation}
\epsilon _{s}=-\frac{\partial \ln R}{\partial \eta }-\frac{1}{2sL}\,\sum_{p}%
\frac{\partial }{\partial \eta }\ln \left[ A_{2}^{2}+A_{3}^{2}+2A_{2}A_{3}%
\cos (\frac{2\pi p}{L})\right] ^{1/2},
\end{equation}
\begin{equation}
D_{\epsilon }(p,\eta )=\frac{1}{\sqrt{1-G^{2}(p,\eta )}}\,\frac{\partial }{%
\partial \eta }\ln G(p,\eta ).
\end{equation}%
Likewise, using the definition given by Eq. (21) we can perform the
derivative of $\epsilon $ to obatin the specific heat $c$ as \ 
\begin{equation}
c=c_{s}+\frac{\eta ^{2}}{2sL}\sum\limits_{p}D_{c}(p,\eta ),  \label{dc}
\end{equation}%
with 
\begin{equation}
c_{s}=\eta ^{2}\frac{\partial ^{2}\ln R}{\partial \eta ^{2}}+\frac{\eta ^{2}%
}{2sL}\,\sum_{p}\frac{\partial ^{2}}{\partial \eta ^{2}}\ln \left[
A_{2}^{2}+A_{3}^{2}+2A_{2}A_{3}\cos (\frac{2\pi p}{L})\right] ^{1/2},
\end{equation}
\begin{equation}
D_{c}(p,\eta )=-\frac{\partial }{\partial \eta }D_{\epsilon }(p,\eta ).
\end{equation}

For the case of periodic boundary condition, our numerical results from the
measurement of specific heat indicate that the shift of critical
temperature, $\delta \left( L\right) $, does obey the scaling form \cite%
{ferdi}, $\delta \left( L\right) =b\ln L/\left( L\right) ^{2}$ with $b=0.135$
for the rectangular lattice, $0.225$ for the triangular lattice, and $0.166$
for the hexagonal lattice when the case of isotropic couplings is
considered. For the case of AJC, the situation, $1-G^{2}(p,\eta )=0$, as
shown in the corollary, occurs at one point, $p=0$ and $\eta =\eta _{c}$.
Then, the left and right derivatives of the free energy are discontinuous at
this location, and this leads to the finite jumps at the critical point.
Thus, the first-order phase transition occurs at $\eta _{c}$ for the system
with AJC. \ \ \ 

\section{Measurement of discontinuity}

To speculate how the finite jumps appear at the critical point, we may
consider, for instance, Eq. (35) for the internal energy. By direct
calculation, one can show that the quantity, $\partial \left[ \ln G(0,\eta )%
\right] /\partial \eta $, vanishes at the critical point $\eta _{c}$. In
this manner, although the denominator, $\sqrt{1-G^{2}(0,\eta _{c})}$, also
vanishes, the value of $D_{\epsilon }(0,\eta )$ evaluated in the limit of
approaching $\eta _{c}$ can be finite. The existence of the jumps is then
related to the sign change of the functions, $\partial \left[ \ln G(0,\eta )%
\right] /\partial \eta $ and $\sqrt{1-G^{2}(0,\eta )}$, across the critical
point. While the function $\partial \left[ \ln G(0,\eta )\right] /\partial
\eta $ appears to be continuous function with nonzero slope intersecting the
real line at the critical point, its value essentially reverses sign across
the critical point. On the other hand, the function $\sqrt{1-G^{2}(0,\eta )}$
must have its minimum at the critical point, and thence it does not change
sign across the critical point. Thus, as approaching the critical point the
left and right limits of Eq. (35) do not coincide for the zeroth mode and
form the jump. Because the zero mode is absent from PJC, the discontinuity
only appears in the case of AJC. \ \ \ 

To measure the discontinuity appearing in the case of AJC, we introduce the
critical gaps, say $\Delta _{\epsilon }$ and $\Delta _{c}$ for the internal
energy and the specific heat respectively, defined as 
\begin{eqnarray}
\Delta _{\epsilon } &=&\lim_{T\rightarrow T_{c}^{+}}\epsilon
(T)-\lim_{T\rightarrow T_{c}^{-}}\epsilon (T)\ ,  \label{cgpi} \\
\Delta _{c} &=&\lim_{T\rightarrow T_{c}^{+}}c(T)-\lim_{T\rightarrow
T_{c}^{-}}c(T).  \label{cgpc}
\end{eqnarray}%
Thus, $\Delta _{\epsilon }$ is the amount of latent heat per site released
from the first-order phase transition. Since only the zero mode contributes
to the discontinuity, according to Eqs. (34) and (36) we have 
\begin{eqnarray}
\Delta _{\epsilon } &=&\frac{1}{2sL}\lim_{\varepsilon \rightarrow 0}\left[
D_{\epsilon }(0,\eta _{c}-\varepsilon )-D_{\epsilon }(0,\eta
_{c}+\varepsilon )\right] , \\
\Delta _{c} &=&\frac{\eta _{c}^{2}}{2sL}\lim_{\varepsilon \rightarrow 0}%
\left[ D_{c}(0,\eta _{c}-\varepsilon )-D_{c}(0,\eta _{c}+\varepsilon )\right]
.
\end{eqnarray}%
Then, we may extract the size-dependence from the critical gaps by
introducing the gap amplitudes, $a_{\epsilon }$ and $a_{c}$, as 
\begin{equation}
\Delta _{\epsilon }=\frac{a_{\epsilon }}{sL}\text{ \ \ and \ \ }\Delta _{c}=%
\frac{a_{c}}{sL}.  \label{gap2}
\end{equation}%
Note that in the above equation we have isolated the the proportionality
factor $s$ between total number of lattice sites $N$ and the length $L$
along the finite edge. Owing to the inversional proportionality to $L$, the
critical gaps in fact vanish in the thermodynamic limit for which the
non-analiticity changes to be the second-order phase transition..

The explicit forms for the gap amplitudes can be expressed as functions of
the coupling ratios $r_{ij}=J_{i}/J_{j}$, and the results read 
\begin{eqnarray}
a_{\epsilon } &=&2+{r_{31}}\left[ {\frac{(1-{t_{3}}^{2})\,(1+{t_{2}}^{2})}{({%
t_{3}}+{t_{2}})\,(1-{t_{2}}\,{t_{3}})}}\right] _{c}+{r_{21}}\left[ {\frac{(1+%
{t_{3}}^{2})\,(1-{t_{2}}^{2})}{({t_{3}}+{t_{2}})\,(1-{t_{2}}\,{t_{3}})}}%
\right] _{c}\text{ }  \label{tGa} \\
\frac{a_{c}}{\left( {\eta _{1}}^{2}\right) _{c}} &=&({r_{31}}-{r_{21}})^{2}%
\left[ {\frac{\,(1-{t_{3}}^{4})\,(1-{t_{2}}^{4})}{({t_{3}}+{t_{2}})^{2}\,(1-{%
t_{2}}\,{t_{3}})^{2}}}\right] _{c}+4{r_{31}}\,{r_{21}}\left[ {\frac{(1-{t_{3}%
}^{2})\,(1-{t_{2}}^{2})}{({t_{3}}+{t_{2}})^{2}}}\right] _{c}
\end{eqnarray}%
for triangular lattice, and 
\begin{eqnarray}
a_{\epsilon } &=&r_{12}\left[ \frac{(1+t_{3}^{2})\,(1-t_{1}^{2})}{%
(t_{3}+t_{1})\,(1-t_{1}\,t_{3})}\right] _{c}+r_{32}\left[ \frac{%
(1-t_{3}^{2})\,(1+t_{1}^{2})}{(t_{3}+t_{1})\,(1-t_{1}\,t_{3})}\right] _{c}+%
\left[ \frac{1-t_{2}^{2}}{t_{2}}\right] _{c}  \label{hG} \\
\frac{a_{c}}{\left( \eta _{2}^{2}\right) _{c}} &=&(r_{32}-r_{12})^{2}\left[ 
\frac{(1-t_{3}^{4})\,(1-t_{1}^{4})}{(t_{3}+t_{1})^{2}\,(1-t_{1}\,t_{3})^{2}}%
\right] _{c}+4r_{32}\,r_{12}\left[ \frac{(1-t_{3}^{2})\,(1-t_{1}^{2})}{%
(t_{3}+t_{1})^{2}}\right] _{c}+\left[ \frac{1-{t_{2}^{4}}}{t_{2}^{2}}\right]
_{c}  \label{hGa}
\end{eqnarray}%
for hexagonal lattice. Here the subscript $c$ indicates that the functions
of temperature contained in the paretheses are evaluated at the critical
point of the Ising transition determined by Eq. (15). Note that the
reference scales for Eqs. (44) and (46) are employed as $J_{1}$ and $J_{2}$
respectively, according to our convention shown in Fig. 1, and the
permutation of the other coupling strengths leave the results invariant in
each case. Again, the results for square lattice can be obtained from Eqs.
(44) and (45) by putting $t_{3}$ and ${r_{31}}$ to zero.

The numerical results of the gap amplitudes for various ratios of coupling
constants are summarized in Tables. I and II. Since the expressions of Eqs.
(44)-(47) all tend to be positive definite, by means of Eqs. (39) and (40),
the left limit for the value of the internal energy or the specific heat at
the critical point is always greater than the right limit. For both the
triangular and hexagonal lattices, the gaps of the internal energy increase
when the coupling constants are strengthened, however for the specific heat
the behaviors appear to be rather complicated. Here, we record solely some
of the typical values subject to the cases of isotropic couplings: For the
triangular lattice $a_{\epsilon }$ and $a_{c}$ are $6.0$ and $0.905212$
respectively, while reducing to the rectangular lattices the values are $4.0$
and $1.098589$. On the other hand, the gap amplitudes for the hexagonal
lattice are subsequently $3.464102$ and $1.734378$ for the internal energy
density and specific heat. \ 

\section{Summary}

Based on the exact study of infinitely long Ising cylinders with the length
of circumference $L$, we show analytically that there exists first-order
phase transition for the case of antiperiodically joined circumference. The
transition point locates at the critical point of the Ising transition. This
peculiar phenomena arises because of the existence of the zero mode of spin
wave. The soft mode of spin waves causes the free energy \ to behave less
smoothly around the critical temperature, and this leads to the finite jumps
in the derivative quantities, such as the internal energy and specific heat
of the system. The finite jumps, referred as the critical gaps, are found to
be inversely proportional to the width of the system. We then extract the
size-dependent factors from the critical gaps by introducing the gap
amplitudes. The magnitudes of the gap amplitudes for the internal energy and
specific heat are also given explicitly in terms of the ratios of the
coupling strengths. In general, the gaps of the internal energy increase
when the coupling constants are strengthened, however for the specific heat
the behaviors appear to be rather complicated. In the limit of $L\rightarrow
\infty $, the nature of the non-analiticity becomes the conventional
two-dimensional Ising transition.

\section{Acknowledgments}

The authors wish to express their gratitude to Prof. V. N. Plechko for
critical reading and constructive comments and to Prof. F. Y. Wu for
stimulated discussions. We also thank Dr. Ming-Chya Wu for useful
discussions and Dr. Wei-Hsiung Chao for fixing some descriptions relevant to
the discontinuity. This work was partially supported by the National Science
Council of Republic of China (Taiwan) under the Grant No. NSC
92-2112-M-033-005.

\newpage 
\begin{table}[hbt]
\caption{A list of some values of the gap amplitudes and the critical
temperatures for the triangular lattice where conventionally $J_{1}$ is used
as the reference scale and the ratios of coupling constants are given as $%
r_{21}=J_{2}/J_{1}$ and $r_{31}=J_{3}/J_{1}$.}
\label{tab:t1}
\begin{center}
\vspace{8pt} $%
\begin{array}{rcccl}
\hline\hline
&  &  &  &  \\ 
r_{21} & r_{31} & {\eta_{1}}_{c}^{-1} & a_{\epsilon^{(1)}}(r_{21},r_{31}) & 
a_{c}(r_{21},r_{31}) \\ \hline
&  &  &  &  \\ 
\frac{1}{16} & \frac{1}{16} & .9749669067 & 2.980303124 & .9945387423 \\ 
\frac{1}{16} & \frac{1}{8} & 1.120609893 & 3.127565685 & .9938210612 \\ 
\frac{1}{16} & \frac{1}{4} & 1.359594319 & 3.357757709 & 1.001236145 \\ 
\frac{1}{16} & \frac{1}{2} & 1.745136904 & 3.697719840 & 1.025135541 \\ 
\frac{1}{16} & 1 & 2.368135472 & 4.172052393 & 1.070033044 \\ 
\frac{1}{16} & 2 & 3.387294494 & 4.794576091 & 1.124851910 \\ 
\frac{1}{16} & 4 & 5.080921897 & 5.561554227 & 1.158588497 \\ 
\frac{1}{16} & 8 & 7.939611600 & 6.453530816 & 1.136952941 \\ 
\frac{1}{16} & 16 & 12.83539772 & 7.444318063 & 1.046706223 \\ 
\frac{1}{8} & \frac{1}{8} & 1.248671356 & 3.265311339 & .9867447193 \\ 
\frac{1}{8} & \frac{1}{4} & 1.471030695 & 3.492162591 & .9857854706 \\ 
\frac{1}{8} & \frac{1}{2} & 1.845058443 & 3.839911133 & 1.002261462 \\ 
\frac{1}{8} & 1 & 2.464588209 & 4.335206961 & 1.044751239 \\ 
\frac{1}{8} & 2 & 3.490273808 & 4.990854252 & 1.104142168 \\ 
\frac{1}{8} & 4 & 5.202643928 & 5.800546669 & 1.147521663 \\ 
\frac{1}{8} & 8 & 8.096230369 & 6.742319380 & 1.136541280 \\ 
\frac{1}{8} & 16 & 13.05035679 & 7.788226528 & 1.054645265 \\ 
\frac{1}{4} & \frac{1}{4} & 1.674907167 & 3.724368061 & .970816527 \\ 
\frac{1}{4} & \frac{1}{2} & 2.034628498 & 4.095394305 & .9707838264 \\ 
\frac{1}{4} & 1 & 2.650888515 & 4.637804574 & 1.003032167 \\ 
\frac{1}{4} & 2 & 3.690116886 & 5.363639576 & 1.065812648 \\ 
\frac{1}{4} & 4 & 5.438377279 & 6.261540830 & 1.124690886 \\ 
\frac{1}{4} & 8 & 8.398210101 & 7.304442724 & 1.133261620 \\ 
\frac{1}{4} & 16 & 13.46287486 & 8.461116081 & 1.067584184 \\ 
\frac{1}{2} & \frac{1}{2} & 2.383122015 & 4.521379708 & .9433175594 \\ 
\frac{1}{2} & 1 & 3.001777420 & 5.165044415 & .9477520819 \\ 
\frac{1}{2} & 2 & 4.069256996 & 6.039616654 & 1.002111403 \\ 
\frac{1}{2} & 4 & 5.884122781 & 7.122162939 & 1.079055707 \\ 
\frac{1}{2} & 8 & 8.964879141 & 8.372913203 & 1.119838839 \\ 
\frac{1}{2} & 16 & 14.23051189 & 9.753480244 & 1.084320289 \\ 
1 & 1 & 3.640956906 & 6.000000000 & .9052117204 \\ 
1 & 2 & 4.766244030 & 7.172866022 & .9199187375 \\ 
1 & 4 & 6.699628670 & 8.639380759 & .9984555121 \\ 
1 & 8 & 9.989370850 & 10.32255458 & 1.079677481 \\ 
1 & 16 & 15.59947050 & 12.16074113 & 1.094445451 \\ 
2 & 2 & 6.003554839 & 8.866690995 & .864293479 \\ 
2 & 4 & 8.138513992 & 11.08035144 & .8938292729 \\ 
2 & 8 & 11.76824556 & 13.65416899 & .9930189587 \\ 
2 & 16 & 17.92975829 & 16.43931998 & 1.071852827 \\ 
4 & 4 & 10.60355406 & 14.54302699 & .8300549508 \\ 
4 & 8 & 14.76046755 & 18.81937049 & .8743124974 \\ 
4 & 16 & 21.75350911 & 23.54209421 & .9878571688 \\ 
8 & 8 & 19.71670567 & 25.86585276 & .8066370472 \\ 
8 & 16 & 27.92219046 & 34.25059835 & .8619605402 \\ 
16 & 16 & 37.89016754 & 48.49746442 & .7926740283 \\ \hline\hline
\end{array}
$%
\end{center}
\end{table}
\begin{table}[hbt]
\caption{A list of some values of the gap amplitudes and the critical
temperatures for the hexagonal lattice where conventionally $J_{2}$ is used
as the reference scale and the ratios of coupling constants are given as $%
r_{12}=J_{1}/J_{2}$ and $r_{32}=J_{3}/J_{2}$.}
\label{tab:h1}
\begin{center}
\vspace{8pt} $%
\begin{array}{rcccl}
\hline\hline
&  &  &  &  \\ 
r_{12} & r_{32} & {\eta_{2}}_{c}^{-1} & a_{\epsilon^{(2)}}(r_{12},r_{32}) & 
a_{c}(r_{12},r_{32}) \\ \hline
&  &  &  &  \\ 
\frac{1}{16} & \frac{1}{16} & .1418239113 & .1767795664 & .7771176184 \\ 
\frac{1}{16} & \frac{1}{8} & .2051068932 & .2410488007 & .8582854546 \\ 
\frac{1}{16} & \frac{1}{4} & .3088086098 & .3121674265 & 1.108064395 \\ 
\frac{1}{16} & \frac{1}{2} & .4686005983 & .4103205597 & 1.546763926 \\ 
\frac{1}{16} & 1 & .6558979475 & .5033239853 & 1.751829109 \\ 
\frac{1}{16} & 2 & .7724015781 & .4723577917 & 1.326421433 \\ 
\frac{1}{16} & 4 & .7882560748 & .4439584846 & 1.045650432 \\ 
\frac{1}{16} & 8 & .7883670985 & .4434415578 & 1.037622919 \\ 
\frac{1}{16} & 16 & .7883671029 & .4434415117 & 1.037621633 \\ 
\frac{1}{8} & \frac{1}{8} & .2832579533 & .3566830745 & .8618580815 \\ 
\frac{1}{8} & \frac{1}{4} & .4052684179 & .5062127293 & 1.194463434 \\ 
\frac{1}{8} & \frac{1}{2} & .5808413289 & .7041227662 & 1.731487056 \\ 
\frac{1}{8} & 1 & .7855971775 & .8719380815 & 1.997314795 \\ 
\frac{1}{8} & 2 & .9372011965 & .8436299806 & 1.632312177 \\ 
\frac{1}{8} & 4 & .9716704807 & .7734965312 & 1.172806220 \\ 
\frac{1}{8} & 8 & .9723443373 & .7699074318 & 1.136392329 \\ 
\frac{1}{8} & 16 & .9723445189 & .7699052866 & 1.136352737 \\ 
\frac{1}{4} & \frac{1}{4} & .5447699531 & .7916267585 & 1.448073250 \\ 
\frac{1}{4} & \frac{1}{2} & .7347791228 & 1.156043202 & 1.804948783 \\ 
\frac{1}{4} & 1 & .9621164771 & 1.469891246 & 2.068392041 \\ 
\frac{1}{4} & 2 & 1.161682658 & 1.498167751 & 1.897602675 \\ 
\frac{1}{4} & 4 & 1.235234439 & 1.350640815 & 1.302590480 \\ 
\frac{1}{4} & 8 & 1.239071584 & 1.329210535 & 1.170089563 \\ 
\frac{1}{4} & 16 & 1.239077752 & 1.329133341 & 1.169219758 \\ 
\frac{1}{2} & \frac{1}{2} & .9422627892 & 1.757887921 & 1.822399379 \\ 
\frac{1}{2} & 1 & 1.202728348 & 2.359919359 & 1.924993360 \\ 
\frac{1}{2} & 2 & 1.469558246 & 2.614813804 & 1.996128066 \\ 
\frac{1}{2} & 4 & 1.621073672 & 2.399820674 & 1.510617828 \\ 
\frac{1}{2} & 8 & 1.640855146 & 2.297555959 & 1.158833094 \\ 
\frac{1}{2} & 16 & 1.641017920 & 2.295597987 & 1.146474198 \\ 
1 & 1 & 1.518651435 & 3.464101615 & 1.734378103 \\ 
1 & 2 & 1.884525578 & 4.345737934 & 1.798424589 \\ 
1 & 4 & 2.179079812 & 4.362763884 & 1.749768675 \\ 
1 & 8 & 2.266063626 & 4.031943754 & 1.200051039 \\ 
1 & 16 & 2.269182580 & 4.000061955 & 1.098975386 \\ 
2 & 2 & 2.405456695 & 6.447419590 & 1.518194397 \\ 
2 & 4 & 2.939116492 & 7.746094625 & 1.657241194 \\ 
2 & 8 & 3.242147344 & 7.376434599 & 1.457687968 \\ 
2 & 16 & 3.281710292 & 7.093170268 & 1.070763279 \\ 
4 & 4 & 3.848465910 & 11.93486884 & 1.320066912 \\ 
4 & 8 & 4.646730633 & 13.85749388 & 1.564461607 \\ 
4 & 16 & 4.940937757 & 13.00580644 & 1.226308460 \\ 
8 & 8 & 6.284777421 & 22.37788617 & 1.177649562 \\ 
8 & 16 & 7.497609571 & 25.18250264 & 1.510718425 \\ 
16 & 16 & 10.49436716 & 42.57519042 & 1.089894205 \\ \hline\hline
\end{array}
$%
\end{center}
\end{table}

\newpage\ Figure1.The planar (a) triangular and (b) hexagonal lattices with
directional coupling constants, $J_{1}$, $J_{2}$, and $J_{3}.$

Figure2.The values of $1-G^{2}\left( p,\eta \right) $ as a function of $\eta 
$ for a given $p$-mode specified in the curve. Here we take $L=6$.

\bigskip

\end{document}